\documentclass[aps,showpacs,preprint]{revtex4}
\usepackage{graphicx}
\usepackage{epsfig}
\DeclareGraphicsExtensions{.eps}

\begin{document}
\title{Spin-aligned neutron-proton pairs in $N=Z$ nuclei}
\author{
S.~Zerguine$^1$
and
P.~Van~Isacker$^2$}
\affiliation{
$^1$Department of Physics, PRIMALAB Laboratory,
University of Batna, Avenue Boukhelouf M El Hadi,
05000 Batna, Algeria}
\affiliation{
$^2$Grand Acc\'el\'erateur National d'Ions Lourds,
CEA/DSM--CNRS/IN2P3, B.P.~55027, F-14076 Caen Cedex 5, France}
\begin {abstract}
A study is carried out of the role of the aligned neutron-proton pair
with angular momentum $J=9$ and isospin $T=0$
in the low-energy spectroscopy
of the $N=Z$ nuclei $^{96}$Cd, $^{94}$Ag, and $^{92}$Pd.
Shell-model wave functions resulting from realistic interactions
are analyzed in terms of a variety of two-nucleon pairs
corresponding to different choices
of their coupled angular momentum $J$ and isospin $T$.
The analysis is performed exactly for four holes ($^{96}$Cd)
and carried further for six and eight holes ($^{94}$Ag and $^{92}$Pd)
by means of a mapping to an appropriate version
of the interacting boson model.
The study allows the identification
of the strengths and deficiencies of the aligned-pair approximation.
\end{abstract}
\maketitle

\section{Introduction}
\label{s_intro}
The study of nuclei with equal numbers of neutrons and protons ($N=Z$)
is one of the declared goals of radioactive-ion-beam facilities,
either in operation or under construction.
Such nuclei are well known when $N$ and $Z$ are small
but, as the atomic mass number $A=N+Z$ increases,
they lie increasingly closer to the proton drip line
and therefore become more difficult to study experimentally.
Nevertheless, several phenomena of interest,
such as the breaking of isospin symmetry
or the emergence of new collective modes of excitation,
are predicted to become more pronounced with increasing $A$,
and this constitutes the main argument for undertaking
the difficult studies of ever more heavier $N=Z$ nuclei.

Arguably, the goal of most interest in this quest
is the uncovering of effects
due to isoscalar ($T=0$) neutron-proton (n-p) pairing.
In contrast to the usual isovector ($T=1$) pairing,
where the orbital angular momenta
and the spins of two nucleons are both antiparallel ({\it i.e.}, $L=0$ and $S=0$),
isoscalar pairing requires the spins of the nucleons to be parallel ($S=1$),
resulting in a total angular momentum $J=1$.
Collective correlation effects are predicted to occur
as a result of isoscalar n-p pairing~\cite{Warner06}
but have resisted so far experimental confirmation because
(i) the $jj$ coupling scheme which is applicable in all but the lightest nuclei
disfavors the formation of isoscalar n-p pairs with $L=0$~\cite{Juillet00},
and (ii) the states associated with this collective mode
are of low angular momentum $J$
and often hidden among high-$J$ isomeric states
which hinders their experimental detection.

Currently, $N=Z$ experiments are approaching $^{100}$Sn,
involving studies of nuclei such as $^{92}$Pd~\cite{Cederwall10}
where nucleons are dominantly confined to the $1g_{9/2}$ orbit.
In the context of these experiments,
Blomqvist recently proposed~\cite{Blomqvist} that
a realistic description of shell-model wave functions
can be obtained in terms of isoscalar n-p pairs
which are completely {\em aligned} in angular momentum,
that is, pairs with $J=9$.
This proposal is attractive
since it is well adapted to the $jj$ coupling scheme,
valid in this mass region,
and because it should encompass the description
of high-$J$ isomeric states.
Blomqvist's idea is related to the so-called stretch scheme
which was advocated a long time ago by Danos and Gillet~\cite{Danos67}.
In the stretch scheme, shell-model states are constructed from aligned n-p pairs,
treated in a quasi-boson approximation
which neglects antisymmetry between the nucleons in different pairs.
The latter approximation is absent from Blomqvist's approach.

In this paper we examine Blomqvist's proposal
for the $N=Z$ nuclei $^{96}$Cd, $^{94}$Ag, and $^{92}$Pd.
We consider several realistic two-body interactions for the $1g_{9/2}$ orbit
and analyze the shell-model wave functions, obtained with these interactions,
of the four-nucleon-hole system ($^{96}$Cd),
in terms of a variety two-pair states.
For the six- and eight-hole nuclei ($^{94}$Ag and $^{92}$Pd)
a direct shell-model analysis in terms of pair states is more difficult,
and we prefer therefore to carry out an indirect check
by means of a mapping to a corresponding boson model.
In these cases our approach is intermediate
between that of Blomqvist~\cite{Blomqvist} and of Danos and Gillet~\cite{Danos67}.
The boson mapping takes care of antisymmetry effects
in an exact manner on the level of four nucleons
but becomes approximate for more.

This paper is organized as follows.
First, some necessary concepts and techniques are introduced:
the formulas needed to carry out
a shell-model calculation in a pair basis
are given in Sect.~\ref{s_4p}
and two mapping techniques
from an interacting fermion to an interacting boson model
are reviewed in Sect.~\ref{s_map}.
The results of our analysis of $N=Z$ nuclei
are presented and discussed in Sect~\ref{s_applic}.
The conclusions and outlook of this work
are summarized in Sect.~\ref{s_conc}.

\section{Four-particle matrix elements}
\label{s_4p}
This section summarizes the necessary ingredients
to carry out calculations in an isospin formalism.
The formulas given are valid for fermions as well as for bosons.
Four-particle states are described
by grouping the particles in two pairs.
These two-pair states can be used, for example, 
as a basis in a shell-model calculation,
facilitating the subsequent analysis
of the pair structure of the eigenstates.
Furthermore, the two-pair representation of four-particle states
is the natural basis to map the shell model
onto a corresponding model in terms of bosons.
Once this mapping is carried out,
the original interacting fermion problem
is reduced to one of interacting bosons,
which can also be solved with the results
summarized in this section.

In the pair representation of four particles
with angular momentum $j$ and isospin $t$
(both integer for bosons and half-odd-integer for fermions),
a state can be written as $|(jt)^2(J_1T_1)(jt)^2(J_2T_2);JT\rangle$
where particles 1 and 2 are coupled to angular momentum and isospin $J_1T_1$,
particles 3 and 4 to $J_2T_2$,
and the intermediate quantum numbers $J_1T_1$ and $J_2T_2$ to total $JT$.
It is convenient to introduce a short-hand notation
that takes care simultaneously
of angular momentum and isospin quantum numbers
and we denote $JT$ as $\Gamma$;
indices will always be carried over consistently,
{\it i.e.}, $\Gamma_i$ refers to $J_iT_i$, $\gamma$ to $jt$, etc.
The state $|(jt)^2(J_1T_1)(jt)^2(J_2T_2);JT\rangle$
is then denoted as $|\gamma^2(\Gamma_1)\gamma^2(\Gamma_2);\Gamma\rangle$.
This state is not (anti)symmetric in all four bosons (fermions)
but can be made so by applying the (anti)symmetrization operator $\hat P$,
\begin{eqnarray}
|\gamma^4[\Gamma_1\Gamma_2]\Gamma\rangle
&\equiv&{\hat P}|\gamma^2(\Gamma_1)\gamma^2(\Gamma_2);\Gamma\rangle
\nonumber\\
&=&\sum_{\Gamma_a\Gamma_b}
[\gamma^2(\Gamma_a)\gamma^2(\Gamma_b);\Gamma|\}\gamma^4[\Gamma_1\Gamma_2]\Gamma]
%\nonumber\\&&\times
|\gamma^2(\Gamma_a)\gamma^2(\Gamma_b);\Gamma\rangle,
\label{e_4pstate}
\end{eqnarray}
where $[\gamma^2(\Gamma_a)\gamma^2(\Gamma_b);\Gamma|\}
\gamma^4[\Gamma_1\Gamma_2]\Gamma]$
is a four-to-two coefficient of fractional parentage (CFP).
The notation in square brackets $[\Gamma_1\Gamma_2]$ implies that~(\ref{e_4pstate})
is constructed from a parent state~\cite{Talmi93}
with intermediate angular momenta and isospins $J_1T_1$ and $J_2T_2$.
It is implicitly assumed
that all pair quantum numbers are allowed,
{\it i.e.} that $J_i$ is even for $T_i=1$ and odd for $T_i=0$.

The four-to-two CFP is given by
\begin{widetext}
\begin{equation}
[\gamma^2(\Gamma_a)\gamma^2(\Gamma_b);\Gamma|\}
\gamma^4[\Gamma_1\Gamma_2]\Gamma]=
\frac{1}{\sqrt{{\cal{N}}^{\Gamma_1\Gamma_2}_{\gamma\Gamma}}}
\left(\delta_{\Gamma_1\Gamma_a}\delta_{\Gamma_2\Gamma_b}+
(-)^{J+T}\delta_{\Gamma_1\Gamma_b}\delta_{\Gamma_2\Gamma_a}
+4\sigma\left[
\begin{array}{ccc}
\gamma & \gamma & \Gamma_a\\
\gamma & \gamma & \Gamma_b\\
\Gamma_1 & \Gamma_2 & \Gamma
\label{e_cfp}
\end{array}
\right]
\right).
\end{equation}
where $\sigma=+1$ for bosons and $\sigma=-1$ for fermions.
Furthermore, ${\cal{N}}^{\Gamma_1\Gamma_2}_{\gamma\Gamma}$ is a normalization constant,
$\delta_{\Gamma_1\Gamma_a}\equiv\delta_{J_1J_a}\delta_{T_1T_a}$, $\delta_{\Gamma_2\Gamma_b}\equiv\delta_{J_2J_b}\delta_{T_2T_b}$, and
\begin{equation}
\left[\begin{array}{ccc}
\gamma & \gamma & \Gamma_a\\
\gamma & \gamma & \Gamma_b\\
\Gamma_1 & \Gamma_2 &\Gamma
\end{array}\right]
=[\Gamma_1][\Gamma_2][\Gamma_a][\Gamma_b]
\left\{\begin{array}{ccc}
j & j & J_a\\
j & j & J_b\\
J_1 & J_2 & J
\end{array}\right\}
\left\{\begin{array}{ccc}
t & t & T_a\\
t & t & T_b\\
T_1 & T_2 & T
\end{array}\right\},
\label{e_ninej}
\end{equation}
\end{widetext}
where the symbol in curly brackets is a nine-$j$ symbol
and $[\Gamma_i]=\sqrt{(2J_i+1)(2T_i+1)}$.
The normalization constant is known in closed form as
\begin{equation}
{\cal{N}}^{\Gamma_1\Gamma_2}_{\gamma\Gamma}=6\left(
1+(-)^{J+T}\delta_{\Gamma_1\Gamma_2}
+4\sigma\left[
\begin{array}{ccc}
\gamma & \gamma & \Gamma_1\\
\gamma & \gamma & \Gamma_2\\
\Gamma_1 & \Gamma_2 & \Gamma
\end{array}
\right]\right).
\label{e_nct}
\end{equation}

The states~(\ref{e_4pstate}) do not form an orthonormal basis.
One needs therefore to determine the overlap matrix elements
which can be written in terms of the CFPs as
\begin{eqnarray}
\langle\gamma^4[\Gamma_1\Gamma_2]\Gamma|
\gamma^4[\Gamma'_1\Gamma'_2]\Gamma\rangle&=&
\sum_{\Gamma_a\Gamma_b}
[\gamma^2(\Gamma_a)\gamma^2(\Gamma_b);\Gamma|\}
\gamma^4[\Gamma_1\Gamma_2]\Gamma]
%\nonumber\\&&\times
[\gamma^2(\Gamma_a)\gamma^2(\Gamma_b);\Gamma|\}
\gamma^4[\Gamma'_1\Gamma'_2]\Gamma]
\nonumber\\&=&
\frac{6}{\sqrt{{\cal{N}}^{\Gamma_1\Gamma_2}_{\gamma\Gamma}}}
[\gamma^2(\Gamma_1)\gamma^2(\Gamma_2);\Gamma|\}
\gamma^4[\Gamma'_1\Gamma'_2]\Gamma].
\label{e_overlap}
\end{eqnarray}
Furthermore, the matrix elements of the two-body part
of the Hamiltonian can be expressed as
\begin{equation}
\langle\gamma^4[\Gamma_1\Gamma_2]\Gamma|\hat H_2
|\gamma^4[\Gamma'_1\Gamma'_2]\Gamma\rangle=
6\sum_{\Gamma_a\Lambda}
[\gamma^2(\Gamma_a)\gamma^2(\Lambda);\Gamma|\}
\gamma^4[\Gamma_1\Gamma_2]\Gamma]
[\gamma^2(\Gamma_a)\gamma^2(\Lambda);\Gamma|\}
\gamma^4[\Gamma'_1\Gamma'_2]\Gamma]
\nu_\Lambda,
\label{e_int}
\end{equation}
where $\nu_\Lambda$ are two-body matrix elements
between normalized two-particle states,
$\nu_\Lambda\equiv\langle\gamma^2;\Lambda|\hat H_2|\gamma^2;\Lambda\rangle$.
The label $\Lambda$ is a short-hand notation
for the two-particles' angular momentum $\lambda$ and isospin $T$.
For example, in the nuclear shell model
the particles are nucleons with half-odd-integer angular momentum $j$
and isospin $t=1/2$.
We recall that in this case
the antisymmetric two-nucleon states
are uniquely determined by the total angular momentum $\lambda$
and that the total isospin is a redundant quantum number
which is $T=0$ for odd $\lambda$ and $T=1$ for even $\lambda$.
The two-body matrix elements
therefore depend on $\lambda$ only,
$\nu_\lambda=\langle j^2;\lambda|\hat H_2|j^2;\lambda\rangle$.
In the interacting boson model, in contrast,
the particles are bosons with integer angular momentum $j$
and integer isospin $t=0$ or $t=1$.
In the latter case, the two-particles' isospin $T$
(obtained by coupling $t=1$ with $t=1$)
is {\em not} redundant
but is needed to fully characterize the two-particle state.

\section{Methods of mapping}
\label{s_map}
The basic and common idea of different boson mapping methods
is to truncate the full shell-model space
to a subspace which is written in terms of fermion pairs
and to establish subsequently a correspondence
between the fermion-pair space
and an analogous space which is written in terms of bosons.
This section recalls briefly two methods,
namely the OAI (Otsuka-Arima-Iachello)~\cite{Otsuka78}
and the democratic~\cite{Skouras90} mappings
that will be used in the applications.

\subsection{The OAI mapping}
\label{s_oai}
We limit ourselves here to the problem of $n$ nucleons
with isospin $t=1/2$ in a single orbit $j$
which is the case of interest in this paper.
A detailed description of the method in its full generality
can be found in Ref.~\cite{Otsuka78}.
 
Let us first introduce the pair creation operators
\begin{equation}
A^{\dag(JT)}_{M_JM_T}=
\frac{1}{\sqrt2}[a^\dag\times a^\dag]^{(JT)}_{M_JM_T},
\label{e_pair}
\end{equation}
where $a^\dag$ is the creation operator of a $t=1/2$ nucleon in orbit $j$
and the square brackets denote coupling to a tensor
with angular momentum $J$ and isospin $T$,
and projections $M_J$ and $M_T$, respectively.
For nucleons in a single orbit $j$,
the pair~(\ref{e_pair}) is totally determined by its angular momentum $J$
and its isospin follows from it
({\it i.e.}, $T=0$ for odd $J$ and $T=1$ for even $J$).
The starting point of the OAI mapping
is a given shell-model Hamiltonian $\hat H^{\rm F}$
which is a scalar in $J$ and $T$,
and the selection of a number of fermion pairs $J_1,J_2,\dots,J_p$,
to each of which will be associated a boson
with the same angular momentum $J_i$.
(For the bosons this angular momentum
is by convention denoted as $\ell_i$, $\ell_i=J_i$.)
For a single-$j$ orbit one may ignore
the single-particle term of the Hamiltonian
which we therefore take to be of pure two-body character
and denote it as $\hat H^{\rm F}_2$.
The one-body term of the mapped boson Hamiltonian $\hat H^{\rm B}$
is directly obtained from the matrix elements
of the shell-model Hamiltonian $\hat H^{\rm F}_2$ between the pair states,
\begin{equation}
\epsilon_{\ell_i}\equiv
\langle \ell_i|\hat H^{\rm B}|\ell_i\rangle=
\langle J_i|\hat H^{\rm F}_2|J_i\rangle.
\label{e_eboson}
\end{equation}
These matrix elements are interpreted as single-boson energies.

The determination of higher-body terms of the mapped boson Hamiltonian
is more complicated,
as we illustrate here for the two-body part.
For a given angular momentum $J$ and isospin $T$,
one first enumerates all possible two-pair states
$|F_i\rangle\equiv|J_aJ_b;JT\rangle$,
where the index $i=1,\dots,d$ corresponds to pairs $(a,b)$ with $1\leq a\leq b\leq p$.
These will be mapped onto the two-boson states
$|B_i\rangle\equiv|\ell_a\ell_b;JT\rangle$,
\begin{equation}
|F_i\rangle\longrightarrow|B_i\rangle,
\quad i=1,\dots,d.
\label{e_corr}
\end{equation}
This correspondence cannot be established directly
since the boson states are orthogonal
while the fermion states are not:
\begin{equation}
\langle B_i|B_j\rangle=\delta_{ij},
\quad
o_{ij}\equiv\langle F_i|F_j\rangle\neq\delta_{ij},
\label{e_orth}
\end{equation}
where $o_{ij}$ is the overlap matrix in fermion space.
To deal with this problem, the strategy of the OAI mapping
is to define for each $JT$ a {\em hierarchy} of pair states,
to apply a Gram-Schmidt {\em orthogonalization} of this ordered sequence,
and to associate the boson states with the {\em orthonormalized} fermion states.
This leads to the definition of the correspondence
\begin{equation}
|\tilde F_i\rangle\Longrightarrow|B_i\rangle,
\quad i=1,\dots,d,
\label{e_oai}
\end{equation}
where the states $|\tilde F_i\rangle$
form an orthonormalized basis.
One has the series
\begin{eqnarray}
|\tilde F_1\rangle&=&\frac{1}{\sqrt{o_{11}}}|F_1\rangle,
\nonumber\\
|\tilde F_2\rangle&=&
\frac{1}{\sqrt{o_{22}-(\tilde o_{21})^2}}
\bigl(|F_2\rangle-\tilde o_{21}|\tilde F_1\rangle\bigr),
\nonumber\\ &\vdots&\nonumber\\
|\tilde F_k\rangle&=&
\frac{1}{\sqrt{{\cal N}_k}}
\left(|F_k\rangle-\sum_{i=1}^{k-1}\tilde o_{ki}|\tilde F_i\rangle\right),
\end{eqnarray}
until $k=p$, with
\begin{equation}
{\cal N}_k=o_{kk}-\sum_{i=1}^{k-1}(\tilde o_{ki})^2,
\quad
\tilde o_{ki}\equiv\langle F_k|\tilde F_i\rangle.
\label{e_gsno}
\end{equation}
An efficient algorithm to carry out the Gram-Schmidt procedure
involves expanding the orthonormal states $|\tilde F_k\rangle$
in the non-orthogonal basis $|F_i\rangle$ as
\begin{equation}
|\tilde F_k\rangle=
\sum_{i=1}^ka_{ki}|F_i\rangle,
\label{e_gsex}
\end{equation}
and calculating the coefficients $a_{ki}$ (needed for $k\geq i$)
recursively from
\begin{equation}
a_{kk}=\frac{1}{\sqrt{{\cal N}_k}},
\qquad
a_{ki}=
-\frac{1}{\sqrt{{\cal N}_k}}
\sum_{i'=i}^{k-1}\tilde o_{ki'}a_{i'i},
\qquad
\tilde o_{ki}=
\sum_{i'=1}^ia_{ii'}o_{ki'},
\label{e_gsa}
\end{equation}
which as only input requires
the knowledge of the overlap matrix elements $o_{ij}$.
The matrix elements of the fermion Hamiltonian
in the orthonormal basis $|\tilde F_k\rangle$
are then given by
\begin{equation}
\langle\tilde F_k|\hat H^{\rm F}_2|\tilde F_l\rangle=
\sum_{ij}a_{ki}a_{lj}\langle F_i|\hat H^{\rm F}_2|F_j\rangle,
\label{e_gsmat}
\end{equation}
and the mapped boson Hamiltonian follows from
\begin{equation}
\langle \ell_a\ell_b;JT|\hat H^{\rm B}|\ell_{a'}\ell_{b'};JT\rangle=
\langle\widetilde{J_aJ_b};JT|\hat H^{\rm F}_2|\widetilde{J_{a'}J_{b'}};JT\rangle.
\label{e_oaiham}
\end{equation}
Note that this equation defines the entire boson Hamiltonian
up to and including two-body interactions.
To isolate its two-body part $\hat H^{\rm B}_2$,
one should subtract the previously determined one-body terms
according to
\begin{equation}
\langle \ell_a\ell_b;JT|\hat H^{\rm B}_2|\ell_{a'}\ell_{b'};JT\rangle=
\langle \ell_a\ell_b;JT|\hat H^{\rm B}|\ell_{a'}\ell_{b'};JT\rangle-
(\epsilon_{\ell_a}+\epsilon_{\ell_b})\delta_{aa'}\delta_{bb'},
\label{e_oaiham2}
\end{equation}
always assuming that $a\leq b$ and $a'\leq b'$.

This is the version of the OAI mapping as it will be applied in this paper.
The boson Hamiltonian is obtained from the two- and four-particle systems
and is kept constant for systems with higher numbers of particles.
This is akin to the democratic mapping
but the latter has the advantage
that no hierarchy of two-pair states is required,
as is discussed in the next subsection.

Nevertheless, the hierarchy imposed in the OAI mapping
is often based on arguments of seniority
which enables an extension to $n$-particle systems.
To illustrate this point,
we consider the case of a truncation onto a subspace
constructed out of pairs of angular momenta $J=0$ ($S$ pair) and $J=2$ ($D$ pair).
(In this example we assume for simplicity identical nucleons
so that isospin can be omitted.)
The creation operators $S^\dag$ and $D^\dag$ are defined as
\begin{equation}
S^\dag=A^{\dag(0)}_0,
\quad
D^\dag_M=A^{\dag(2)}_M.
\label{e_sd1}
\end{equation}
The truncated shell-model space
is the $SD$ subspace spanned by the states
\begin{equation}
|j^nv\xi;JM_J\rangle=
\frac{1}{{\cal N}_{\rm F}}
{\cal P}(S^\dag)^{(n-v)/2}[(D^\dag)^{v/2}]^{(J)}_{M_J}|{\rm o}\rangle, 
\label{e_sdn}
\end{equation}
where $|{\rm o}\rangle$ refers to the closed core,
$\xi$ denotes additional quantum numbers
related to the intermediate angular-momentum couplings of the $D$ pairs,
and ${\cal N}_{\rm F}$ is a normalization constant.
The operator ${\cal P}$ is needed to ensure orthogonality of the basis.
Its effect is a projection onto a subspace with seniority $v$,
the number of particles not in pairs coupled to $J=0$.
As a result the states~(\ref{e_sdn}) are normalized
and can be mapped onto $sd$ states.
This can be expressed by the general correspondence
\begin{equation}
|j^nv\xi;JM_J\rangle\Longrightarrow
|n_sn_d\xi;JM_J\rangle=
\frac{1}{{\cal N}_{\rm B}}(s^\dag)^{n_s}[(d^\dag)^{n_d}]^{(J)}_{M_J}|{\rm o}),
\label{e_sdmap}
\end{equation}
where $|{\rm o})$ is the boson vacuum,
$n_s$ and $n_d$ are the $s$- and $d$-boson numbers, respectively,
with $n_s=(n-v)/2$ and $n_d=v/2$,
and ${\cal N}_{\rm B}$ is a normalization constant.
The matrix elements of the mapped boson Hamiltonian
are now defined as
\begin{equation}
\langle n_sn_d\xi;JM_J|\hat H^{\rm B}|n'_sn'_d\xi';JM_J\rangle=
\langle j^nv\xi;JM_J|\hat H^{\rm F}_2|j^nv'\xi';JM_J\rangle.
\label{e_oaihamn}
\end{equation}
The difference with Eq.~(\ref{e_oaiham})
is that the latter equation is specific to $n=4$ particles
while the mapping (\ref{e_oaihamn}) applies to any $n$.
The calculation of an $n$-particle fermion matrix element is possible
because the seniority formalism allows it to be reduced to $n=\max(v,v')$~\cite{Talmi93}.
Hence, as long as $v,v'\leq4$
%which corresponds to $n_d,n'_d\leq2$,
the fermion matrix element can be computed
and fixes the corresponding boson matrix element.
The advantage of this procedure is
that it determines a boson Hamiltonian
which varies with particle number
yielding a more adequate description of Pauli effects.
Its disadvantage is that a generalization
towards any choice of pairs and/or to neutrons and protons
is difficult, if not impossible.
Therefore, we opt in this paper for the simpler OAI mapping
that defines a constant boson Hamiltonian
from the two- and four-particle systems.

\subsection{The democratic mapping}
\label{s_dem}
Again the starting point is the selection
of a number of fermion pairs $J_1,J_2,\dots,J_p$
and a corresponding series of bosons with energies $\epsilon_{\ell_i}$
determined from Eq.~(\ref{e_eboson}).
To establish the correspondence~(\ref{e_corr}),
complicated by the non-orthogonality of the fermion two-pair states,
the democratic mapping relies on the diagonalization
of the overlap matrix $o_{ij}$ defined in Eq.~(\ref{e_orth}).
This diagonalization provides, besides the eigenvalues $o_k$,
an orthogonal basis $|X_k\rangle$,
\begin{equation}
|X_k\rangle=\sum_ic_{ki}|F_i\rangle,
\quad
k=1,\dots,d,
\label{e_xk}
\end{equation}
with
\begin{equation}
\langle X_k|X_l\rangle=
\sum_{ij}c_{ki}c_{lj}o_{ij}=
o_k\delta_{kl},
\quad
k,l=1,\dots,d.
\label{e_xkxl}
\end{equation}
We follow the convention of labeling non-orthogonal basis states by $i$ or $j$,
and orthogonal basis states by $k$ or $l$. 
The coefficients $c_{ki}\equiv\langle F_i|X_k\rangle\equiv\langle X_k|F_i\rangle$
are transformation coefficients from the non-orthonormal basis $|F_i\rangle$
to the basis $|X_k\rangle$ and they satisfy
\begin{equation}
\sum_ic_{ki}c_{li}=\delta_{kl},
\quad
k,l=1,\dots,d.
\label{e_ckl}
\end{equation}
The states $|X_k\rangle$ form an orthogonal but non-normalized
basis since they satisfy~(\ref{e_xkxl}).
To normalize these states, we define
\begin{equation}
|\bar F_k\rangle=\frac{1}{\sqrt{o_k}}|X_k\rangle,
\quad
k=1,\dots,d,
\label{e_xfk}
\end{equation}
which indeed satisfy
\begin{equation}
\langle\bar F_k|\bar F_l\rangle=\delta_{kl},
\quad
k,l=1,\dots,d.
\label{e_xfkxfl}
\end{equation}
The basis $|\bar F_k\rangle$ has the same role
as the basis $|\tilde F_k\rangle$ in the OAI mapping
but it is important to realize that both bases are {\em not} identical
and lead to different boson Hamiltonians.
The democratic mapping relies
on the definition of a transformation in boson space
which is analogous to the one in fermion space
\begin{equation}
|\bar B_k\rangle=\sum_ic_{ki}|B_i\rangle,
\quad
k=1,\dots,d.
\label{e_xbk}
\end{equation}
From the orthogonality of the basis $|B_i\rangle$
and the properties of the coefficients $c_{ki}$
it can be shown that these states form an orthonormal set,
\begin{equation}
\langle\bar B_k|\bar B_l\rangle=\delta_{kl},
\quad
k,l=1,\dots,d.
\label{e_xbkxbl}
\end{equation}
We have now arrived at fermion and boson bases
that are both orthonormal,
and we can therefore establish the mapping 
\begin{equation}
|\bar F_k\rangle\Longrightarrow|\bar B_k\rangle,
\quad
k=1,\dots,d,
\label{e_dem}
\end{equation}
and determine the matrix elements of the boson Hamiltonian
in this basis,
\begin{equation}
\langle\bar B_k|\hat H^{\rm B}|\bar B_l\rangle=
\langle\bar F_k|\hat H^{\rm F}_2|\bar F_l\rangle,
\quad
k,l=1,\dots,d.
\label{e_demham}
\end{equation}
With use of the inverse of the relation~(\ref{e_xbk}),
of the equality~(\ref{e_demham}), and of Eqs.~(\ref{e_xk}) and (\ref{e_xfk}),
the matrix elements of the boson Hamiltonian in the original basis
can be written in terms of those of the fermion Hamiltonian
(also in the original basis) as
\begin{equation}
\langle B_i|\hat H^{\rm B}|B_j\rangle=
\sum_{kl}\sum_{i'j'}
\frac{1}{\sqrt{o_ko_l}}
c_{ki}c_{ki'}c_{lj}c_{lj'}H^{\rm F}_{i'j'},
\label{e_demham2}
\end{equation} 
where $H^{\rm F}_{i'j'}\equiv\langle F_{i'}|\hat H^{\rm F}_2|F_{j'}\rangle$.
Again, as with the OAI mapping,
this defines the entire boson Hamiltonian
from which the two-body part can be isolated
by applying Eq.~(\ref{e_oaiham2}).

In some cases, $m$ fermion vectors $|F_i\rangle$ are linearly dependent
on the $d-m$ others,
leading to $m$ vanishing eigenvalues of the overlap matrix,
\begin{equation} 
\label{pi_0}
o_i=0,
\quad
i=1,\dots,m.
\end{equation}
This problem can be solved
by excluding from the fermion space $m$ states $|F_i\rangle$
and by calculating the matrix elements of $\hat H^{\rm B}$ from the remaining $d-m$ states.
All other matrix elements of $\hat H^{\rm B}$ are defined such that
\begin{eqnarray}
\langle B_i|\hat H^{\rm B}|B_i\rangle&=&\infty,
\quad
i=d-m+1,\dots,d,
\nonumber\\
\langle B_i|\hat H^{\rm B}|B_j\rangle&=&
\langle B_j|\hat H^{\rm B}|B_i\rangle=0,
\quad
i\neq j,
\label{e_lin}
\end{eqnarray}
where $i$ and $j$ in the last equation
take the values $i=1,\dots,d$ and $j=d-m+1,\dots,d$.
One has still to decide which $m$ vectors
to remove from the fermion space.
To avoid arbitrariness, one chooses the $m$ fermion states
with the smallest overlap with the shell-model states.

\section{Application to $N=Z$ nuclei}
\label{s_applic}
Of particular interest in this work are $N=Z$ nuclei
with neutrons and protons in the $1g_{9/2}$ orbit.
Recently, Blomqvist~\cite{Blomqvist} has conjectured
that a valid interpretation of yrast states in these nuclei
can be obtained in terms of n-p $T=0$ pairs
which are coupled to maximum angular momentum $J=9$
and which therefore can be termed aligned n-p pairs.
This is contrary to the usual interpretation of such states
which involves low-spin pairs with isospin $T=1$ and possibly also with $T=0$.
In this section we examine Blomqvist's proposal
with specific reference to the nuclei $^{96}$Cd, $^{94}$Ag, and $^{92}$Pd,
corresponding to four, six, and eight holes
with respect to the $^{100}$Sn core, respectively.
Our study consists of two separate parts:
the analysis of shell-model wave functions of $^{96}$Cd
in terms of a variety of two-fermion pairs
and the mapping of shell-model onto corresponding boson states
for $^{96}$Cd and $^{92}$Pd,
following the formalism developed in the previous sections.
First, an appropriate shell-model interaction should be determined.

\subsection{Shell-model interaction}
\label{ss_smi}
For the purpose of checking the stability of our results
and the reliability of our conclusions,
we have carried out the analysis for three different shell-model interactions.
The SLGT0 interaction is taken from Serduke {\it et al.}~\cite{Serduke76}.
It was used in the more recent analysis of Herndl and Brown~\cite{Herndl97}
where it was found to give satisfactory results
for the neutron-deficient nuclei in the mass region $A=86$ to 100,
which are of interest in the present study.
A second shell-model interaction,
which shall be named GF,
is taken from Gross and Frenkel~\cite{Gross76}.
Both the SLGT0 and GF interactions are defined
in the $2p_{1/2}+1g_{9/2}$ shell-model space
and, to carry out an analysis in terms of aligned pairs,
it is necessary to renormalize them to the $1g_{9/2}$ orbit.
The resulting two-body matrix elements
are shown in columns 2 and 3 of Table~\ref{t_me}.
The renormalization only affects the
$(\lambda,T)=(0,1)$ and $(1,0)$ matrix elements
since these are the only ones
that also occur in the $2p_{1/2}$ orbit.
\begin{table}
\caption{
Two-body matrix elements in the $1g_{9/2}$ orbit,
in units of MeV,
derived from the interactions SLGT0 and GF,
and from the experimental spectrum of $^{90}$Nb.}
\label{t_me}
\centering
\begin{ruledtabular}
\begin{tabular}{@{}ccrcrcr}
&&\multicolumn{5}{c}
{$\nu_\lambda\equiv\langle (1g_{9/2})^2;\lambda T
|\hat H^{\rm F}_2|(1g_{9/2})^2;\lambda T\rangle$}\\
\cline{3-7}
$(\lambda,T)$&~~~~~~&SLGT0&~~~~~~~&GF&~~~~~~~&Nb90\\
\hline
$(0,1)$&&$-2.392$&&$-2.321$&&$-1.758$\\
$(1,0)$&&$-1.546$&&$-1.524$&&$-1.225$\\
$(2,1)$&&$-0.906$&&$-0.937$&&$-0.573$\\
$(3,0)$&&$-0.747$&&$-0.700$&&$-0.521$\\
$(4,1)$&&$-0.106$&&$-0.160$&&$ 0.064$\\
$(5,0)$&&$-0.423$&&$-0.447$&&$-0.332$\\
$(6,1)$&&$ 0.190$&&$ 0.140$&&$ 0.266$\\
$(7,0)$&&$-0.648$&&$-0.640$&&$-0.481$\\
$(8,1)$&&$ 0.321$&&$ 0.241$&&$ 0.334$\\
$(9,0)$&&$-1.504$&&$-1.752$&&$-1.376$\\
\end{tabular}
\end{ruledtabular}
\end{table}

To avoid the renormalization procedure,
we define a third interaction directly for the $1g_{9/2}$ orbit.
This is done in the following way.
The spectrum of $^{90}$Nb is well known~\cite{NNDC}
and enables the determination
of the particle-hole interaction matrix elements
\begin{equation}
\nu_\lambda^{\rm ph}\equiv
\langle 1g_{9/2}(1g_{9/2})^{-1};\lambda T
|\hat H^{\rm F}_2
|1g_{9/2}(1g_{9/2})^{-1};\lambda T\rangle.
\label{e_vph}
\end{equation}
The isospin $T$ is determined from $\lambda$, that is,
$T=4$ for all states except for $\lambda=0$ which has $T=5$.
The absolute value of the matrix element~(\ref{e_vph}) for the ground state ($\lambda=8$)
is obtained from the binding energies of the surrounding nuclei as
\begin{eqnarray}
\nu_8^{\rm ph}&=&
-\left(E(^{90}{\rm Zr})+E(^{90}{\rm Nb})-E(^{91}{\rm Nb})-E(^{89}{\rm Zr})\right)
\nonumber\\
&=&-783.794-776.895+788.942+771.825
\nonumber\\
&=&0.078~{\rm MeV},
\end{eqnarray}
where $E$ stands for the nuclear binding energy,
taken from the 2003 atomic mass evaluation~\cite{Audi03}
and corrected for the electrons' binding energy~\cite{Lunney03}.
(The minus sign is needed to convert
from binding energy to interaction energy.)
All levels $\lambda=0,\dots,9$ of the particle-hole multiplet~(\ref{e_vph})
are known in $^{90}$Nb and fix the differences $\nu_\lambda^{\rm ph}-\nu_8^{\rm ph}$.
A problem with this procedure
concerns the choice of the relevant $J^\pi=1^+$ state in $^{90}$Nb.
Several of them are observed at low excitation energy
and only one should be taken as a member of the particle-hole multiplet.
We have taken the $J^\pi=1^+$ level at 2.126~MeV
because this state is strongly populated
in the $\beta$-decay of $^{90}$Mo ($\log ft=4.9$)~\cite{Porquet}.
Furthermore, it appears to be the only $1^+$ state observed in the triton spectrum
obtained in the charge-exchange reaction
$^{90}{\rm Zr}(^3{\rm He},t)^{90}{\rm Nb}$~\cite{Fields82}.

Once the particle-hole matrix elements $\nu_\lambda^{\rm ph}$
are derived in this way,
the particle-particle (or hole-hole) matrix elements $\nu_\lambda^{\rm pp}$
are obtained from an inverted Pandya tranformation~\cite{Pandya56}.
Since absolute (as opposed to relative) matrix elements
have been extracted from binding energies,
care should be taken to use the appropriate Pandya transformation.
Equation~(18.63) of Ref.~\cite{Talmi93}
gives the following relation between absolute matrix elements:
\begin{equation}
\nu_\lambda^{\rm ph}=
E_0-\sum_{\lambda'}(2\lambda'+1)
\Bigl\{\begin{array}{ccc}
j&j&\lambda\\
j&j&\lambda'
\end{array}\Bigr\}
\nu_{\lambda'}^{\rm pp},
\label{e_pptoph}
\end{equation}
where $E_0$ is a constant ({\it i.e.}, $\lambda$-independent) interaction energy given by
\begin{equation}
E_0=
\frac{1}{2j+1}\sum_{{\rm all}\;\lambda'}(2\lambda'+1)\nu_{\lambda'}^{\rm pp}+
\frac{2j-1}{2j+1}\sum_{{\rm even}\;\lambda'}(2\lambda'+1)\nu_{\lambda'}^{\rm pp}.
%\frac{1}{2}\sum_{\lambda'}(2\lambda'+1)
%\left[1+(-)^{\lambda'}\frac{2j-1}{2j+1}\right]\nu_{\lambda'}^{\rm pp}.
\label{e_e0pp}
\end{equation}
To express the particle-particle in terms of the particle-hole matrix elements,
as is needed here,
the relation~(\ref{e_pptoph}) can now be inverted in the usual manner,
leading to
\begin{equation}
\nu_\lambda^{\rm pp}=
E_0-\sum_{\lambda'}(2\lambda'+1)
\Bigl\{\begin{array}{ccc}
j&j&\lambda\\
j&j&\lambda'
\end{array}\Bigr\}
\nu_{\lambda'}^{\rm ph}.
\label{e_phtopp}
\end{equation}
The problem with this relation is that
it is of no use as long as the constant $E_0$
is expressed in terms of particle-particle matrix elements,
as in Eq.~(\ref{e_e0pp}).
We need an equation for $E_0$
in terms of particle-hole matrix elements.
This can be obtained
by inserting the expression~(\ref{e_phtopp}) for $\nu_\lambda^{\rm pp}$ in Eq.~(\ref{e_e0pp})
and solving for $E_0$, leading to
\begin{equation}
E_0=
\frac{1}{2j(2j+1)}\sum_{{\rm all}\;\lambda'}(2\lambda'+1)\nu_{\lambda'}^{\rm ph}
-\frac{2j-1}{2j(2j+1)}\nu_0^{\rm ph}.
\label{e_e0ph}
\end{equation}

The resulting particle-particle matrix elements
are shown in column 4 of Table~\ref{t_me} under `Nb90',
where the `pp' index is omitted,
$\nu_\lambda\equiv\nu_\lambda^{\rm pp}$,
as will be done from now on.
Note that the matrix elements thus obtained
differ by a constant from those of Sorlin and Porquet
(see Fig.~6 of Ref.~\cite{Sorlin08}),
since we have taken care here of the constant interaction energy $E_0$.

It is seen from Table~\ref{t_me}
that the matrix elements of the SLGT0 and GF interactions are similar.
The biggest difference concerns the $(\lambda,T)=(9,0)$ matrix element
which is more attractive by $\sim250$~keV for GF.
This conceivably might influence
the approximation in terms of aligned $\lambda=9$ n-p pairs.
The interaction derived from $^{90}$Nb is less attractive (or more repulsive),
in particular the pairing matrix element with $(\lambda,T)=(0,1)$.
All three interactions, while having reasonable characteristics,
are sufficiently different to test the robustness of our analysis.

Before proceeding with the wave-function analysis,
we first check to what extent the $N=Z$ nuclei in this region
can be described by confining nucleons to the $1g_{9/2}$ orbit.
This approximation should be reasonable for nuclei
close to (south-west of) $^{100}$Sn
but will become increasingly poor as one approaches $^{80}$Zr.
The latter nucleus is known to be deformed~\cite{Lister87}
and hence a single (spherical) orbit will not suffice 
for a reliable description of nuclei in its neighborhood.
To quantify the limitation to the $1g_{9/2}$ orbit,
we show in Table~\ref{t_p1g9} the results of
a shell-model calculation with the GF interaction.
The table shows the fractions of  $2p_{1/2}+1g_{9/2}$ shell-model eigenstates
of $^{96}$Cd and $^{92}$Pd that lie within the $1g_{9/2}$ subspace,
expressed in percentages.
It is seen that for the lowest eigenstates this fraction is large,
hence justifying the restriction to the $1g_{9/2}$ orbit.
Therefore, we henceforth restrict the shell-model space to $1g_{9/2}$,
in which case the formalism of Sect.~\ref{s_4p} can be applied.
\begin{table}
\caption{
Fractions of  $2p_{1/2}+1g_{9/2}$ shell-model eigenstates of the GF interaction
%of $^{96}$Cd and $^{92}$Pd
that lie within the $1g_{9/2}$ subspace,
expressed in percentages.}
\label{t_p1g9}
\begin{ruledtabular}
\begin{tabular}{@{}cccccccccc}
$J^\pi_i$&$0^+_1$&$0^+_2$&$2^+_1$&$2^+_2$&$4^+_1$&$4^+_2$&$6^+_1$&$8^+_1$&$10^+_1$\\
\hline
$^{96}$Cd&96&86&98&97&98&98&98&97&99\\
$^{92}$Pd&90&95&92&95&94&93&95&94&96\\
\end{tabular}
\end{ruledtabular}
\end{table}

\subsection{Shell-model analysis of the $(1g_{9/2})^4$ system}
\label{ss_sma}
With the formalism developed in Sect.~\ref{s_4p}
it is now possible to perform a shell-model calculation
for four holes in the $1g_{9/2}$ orbit
and analyze the resulting wave functions
in terms of two-pair states.
We concentrate on the $N=Z$ case
which corresponds to two neutrons and two protons.
For a given angular momentum $J$,
the overlap matrix between all possible two-pair states $|(jt)^4[(J_1T_1)(J_2T_2)]JT\rangle$
is constructed using Eq.~(\ref{e_overlap}).
The number $d$ of linearly independent states $JT$
is given by the number of non-zero eigenvalues of the overlap matrix.
We then select $d$ linearly independent but otherwise arbitrary
two-pair states that shall be denoted in short as $|F_i\rangle,i=1,\dots,d$.
Diagonalization of the overlap matrix $o_{ij}\equiv\langle F_i|F_j\rangle$
allows the definition of an orthonormal basis
\begin{equation}
|\bar F_k\rangle=
\sum_i\bar c_{ki}|F_i\rangle,
\quad
k=1,\dots,d,
\label{e_xfk2}
\end{equation}
with $\bar c_{ki}=c_{ki}/\sqrt{o_k}$ in the notation of Sect.~\ref{s_map}.
This basis can now be used to construct the energy matrix,
\begin{equation}
\langle\bar F_k|\hat H^{\rm F}_2|\bar F_l\rangle
=\sum_{ij}\bar c_{ki}\bar c_{lj}\langle F_i|\hat H^{\rm F}_2|F_j\rangle,
%\quad k,l=1,\dots,d,
\label{e_xfk3}
\end{equation}
to be computed with use of Eq.~(\ref{e_int}).
The diagonalization of this matrix
leads to the energy eigenvalues $\epsilon_r,r=1,\dots,d$,
with corresponding eigenvectors given by
\begin{equation}
|E_r\rangle
=\sum_ke_{rk}|\bar F_k\rangle
=\sum_{ki}e_{rk}\bar c_{ki}|F_i\rangle.
%\quad r=1,\dots,d.
\label{e_eig}
\end{equation}
Note that the energies $\epsilon_r$ are independent
of the choice of two-pair states $|F_i\rangle$
as long as the latter span the entire $JT$ space.
Although the above procedure might seem rather cumbersome
for performing a four-hole shell-model calculation,
it has the advantage of providing us directly with the pair structure
of the shell-model eigenstates from the overlaps
\begin{equation}
\langle F_i|E_r\rangle
=\sum_{kj}e_{rk}\bar c_{kj}o_{ij}.
%\quad r=1,\dots,d.
\label{e_overlapsm}
\end{equation}

\begin{figure}
\includegraphics[width=8.5cm]{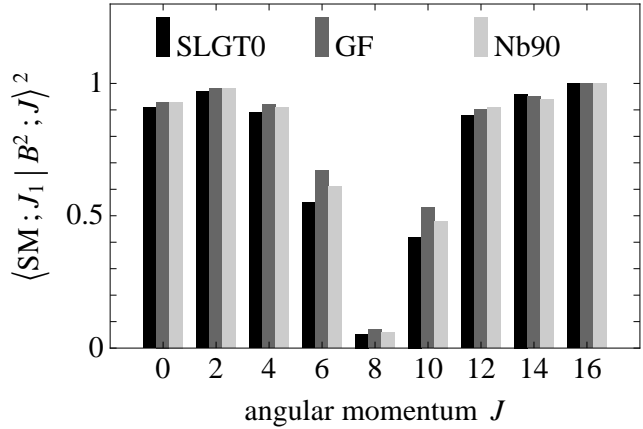}
\caption{
\label{f_bb}
Overlaps of the $(1g_{9/2})^4$ yrast eigenstates
of the interactions SLGT0, GF, and Nb90
with angular momentum $J$ and isospin $T=0$,
with the two-pair state $|B^2;J\rangle$.}
\end{figure}
We now analyze the yrast eigenstates
of the various interactions defined in Table~\ref{t_me},
that is, the quantities $\langle F_i|E_r\rangle^2$
for $r=1$ and a variety of two-pair states $F_i$.
Pairs with angular momentum $J$
are generically denoted as $P_J$
and explicitly, following standard spectroscopic notation,
as $S$, $D$, $G$, $I$, and $K$ for $J=0$, 2, 4, 6, and 8.
Since this paper deals in particular with the aligned n-p pair with $J=9$,
we reserve for it the non-standard notation `$B$' (for Blomqvist).
The central result is shown in Fig.~\ref{f_bb}
which displays the quantity $\langle J_1|B^2;J\rangle^2$
where $|J_1\rangle$ is the yrast eigenstate
with angular momentum $J$ and isospin $T=0$
of the interactions SLGT0, GF, and Nb90.
Most yrast states have a large overlap with $B^2$,
as was shown by Blomqvist,
but this is conspicuously not the case for $J\approx8$.
It seems as if the two aligned n-p pairs
do not like to couple to a total angular momentum
which equals their individual spins.

\begin{table*}
\caption{
Coefficients $a_{[\Gamma_1\Gamma_2]\Gamma}^\lambda$
in the expansion~(\ref{e_dia}) of the diagonal energies
of the $(9/2)^4$ system for isospin $T=0$.}
\label{t_dia}
\begin{ruledtabular}
\begin{tabular}{ccccccccccccccccccc}
&&\multicolumn{2}{c}{$J=0$}&&\multicolumn{2}{c}{$J=2$}
&&\multicolumn{2}{c}{$J=4$}&&\multicolumn{2}{c}{$J=6$}
&&\multicolumn{2}{c}{$J=8$}&&\multicolumn{2}{c}{$J=10$}\\
\cline{3-4}\cline{6-7}\cline{9-10}\cline{12-13}\cline{15-16}\cline{18-19}
$\lambda$&&
$[B^2]$&$[S^2]$&&$[B^2]$&$[SD]$&&$[B^2]$&$[SG]$&&
$[B^2]$&$[SI]$&&$[B^2]$&$[SK]$&&$[B^2]$&$[DK]$\\
\hline
0&&1.14&2.20&&0.85&1.20&&0.42&1.20&&0.11&1.20&&0.01&1.20&&---&---\\
1&&0.76&0.16&&0.64&0.28&&0.42&0.24&&0.18&0.17&&0.04&0.08&&0.00&0.02\\
2&&1.70&0.09&&1.79&1.30&&1.74&0.08&&1.26&0.04&&0.55&0.07&&0.10&1.15\\
3&&0.23&0.38&&0.33&0.47&&0.50&0.21&&0.57&0.28&&0.43&0.52&&0.18&0.11\\
4&&0.16&0.16&&0.34&0.15&&0.79&1.26&&1.41&0.21&&1.76&0.08&&1.36&0.10\\
5&&0.01&0.60&&0.02&0.39&&0.09&0.54&&0.24&0.62&&0.47&0.68&&0.62&0.66\\
6&&0.00&0.24&&0.00&0.11&&0.05&0.30&&0.21&1.29&&0.65&0.20&&1.42&0.21\\
7&&0.00&0.82&&0.00&0.39&&0.00&0.94&&0.01&0.97&&0.06&0.77&&0.19&1.18\\
8&&0.00&0.31&&0.00&0.23&&0.00&0.16&&0.00&0.26&&0.02&1.45&&0.12&1.53\\
9&&2.00&1.04&&2.00&1.47&&2.00&1.07&&2.00&0.96&&2.00&0.95&&2.00&1.03\\
\end{tabular}
\end{ruledtabular}
\end{table*}
To acquire some insight in this finding,
we note that simple expressions
in terms of the two-body matrix elements
are available for the {\em diagonal} energies
of two-pair states,
\begin{equation}
E(\gamma^4[\Gamma_1\Gamma_2]\Gamma)\equiv
\langle\gamma^4[\Gamma_1\Gamma_2]\Gamma|
\hat H^{\rm F}_2
|\gamma^4[\Gamma_1\Gamma_2]\Gamma\rangle=
\sum_\lambda
a_{[\Gamma_1\Gamma_2]\Gamma}^\lambda\;\nu_\lambda,
\label{e_dia}
\end{equation}
where the coefficients $a_{[\Gamma_1\Gamma_2]\Gamma}^\lambda$
of relevance to the present discussion
are shown in Table~\ref{t_dia}.
According to a recent paper by Talmi~\cite{Talmi10},
they are non-negative rational numbers;
since they involve ratios of rather large integers in this case,
Table~\ref{t_dia} gives numerical approximations. 
Of all the coefficients given in this table,
the important ones have $\lambda=0$, 1, and 9,
because these are the multipolarities
of the most attractive interaction matrix elements.
It is seen that the contributions
of the aligned matrix element ($\lambda=9$)
to the energies of the states
$|(9/2)^4[B^2]J\rangle$ and $|(9/2)^4[SP_J]J\rangle$
remain more or less constant, independent of $J$.
This is not the case for the pairing matrix element ($\lambda=0$)
whose contribution to the energy of $|(9/2)^4[B^2]J\rangle$
disappears as $J$ increases
while it remains important for $|(9/2)^4[SP_J]J\rangle$.
The combined effect of these contributions
is that the state $|(9/2)^4[SP_J]J\rangle$
dips below $|(9/2)^4[B^2]J\rangle$ around $J\approx8$
and as a result picks up the largest component of the yrast eigenstate.
Hence, the feature of the disappearing $B$ dominance around $J\approx8$
is explained by a combination of geometry---the CFPs in the $j=9/2$ orbit,
and dynamics---the dependence of the interaction matrix elements on $\lambda$.

\begin{table}
\caption{Overlaps of  the $(1g_{9/2})^4$ yrast eigenstates of the SLGT0 interaction
with angular momentum $J$ and isospin $T=0$ 
with various two-pair states, expressed in percentages.}
\label{t_g9}
\centering
\begin{ruledtabular}
\begin{tabular}{@{}rrrrrrrrrr}
$J$&$B^2$&$SP_J$&$D^2$&$DG$&$DI$&$DK$&$G^2$&$I^2$&$K^2$\\
\hline
$0$  &91     &80         &35     &---     &---    &---      &18     &7.4   &1.9\\
$2$  &97     &85         &17     &22     &---    &---      &1.5    &0.0   &0.4\\
$4$  &89     &64         &42     &11     &11    &---      &0.2    &0.2   &0.0\\
$6$  &55     &70         &---     &43     &0.2   &4.3     &0.0    &0.2   &0.0\\
$8$  &5.3    &83         &---     &---      &7.4   &24      &1.8    &0.2   &0.1\\
$10$&42     &---         &---     &---      &---    &58       &---     &6.1   &0.5\\
$12$&88     &---         &---     &---      &---    &---       &---     &57   &1.5\\
$14$&96     &---         &---     &---      &---    &---       &---     &---   &31.4\\
$16$&100   &---         &---     &---      &---    &---       &---     &---   &100\\
\end{tabular}
\end{ruledtabular}
\end{table}
Figure~\ref{f_bb} shows
that the overlaps $\langle J_1|B^2;J\rangle^2$
are very similar for the three interactions.
This finding is at the basis of the fact
that the subsequent analysis
gives consistent results for the three interactions.
While there can be significant differences
in the shell-model results with the different interactions,
the approximation in terms of aligned pairs
is similar for the three interactions.
In other words, if a particular shell-model state
is well approximated in terms of aligned pairs for one interaction,
it is so for the other two as well;
if the approximation is less good,
it is so for all three interactions.
Although we have carried out the complete analysis
for the three interactions SLGT0, GF, and Nb90,
we will show in the following only the results of the former
since it has a proven track record of satisfactorily reproducing
the data in the mass region of interest~\cite{Herndl97}.

In Table~\ref{t_g9} are shown the amplitudes in percentages
for yrast eigenstates of the SLGT0 interaction with even $J$ and $T=0$,
that is, the quantities $100\times\langle F_i|E_r\rangle^2$
for $r=1$ and a variety of pair states $|F_i\rangle$.
The numbers illustrate that,
at least at low and at high angular momentum $J$,
the overlaps of the physical eigenstates with $|B^2;J\rangle$
are more important than those with other pair combinations.
The percentages shown in Table~\ref{t_g9}
also illustrate the non-orthogonality of the two-pair basis.
For the example, the $J=0$ ground state has a 91\% overlap with $B^2$
but also a 80\% overlap with $S^2$;
this can only be if the overlap $\langle B^2|S^2\rangle$ itself
is rather large.

\subsection{Boson mappings}
\label{ss_map}
Ideally, one would like to perform a similar analysis
of shell-model eigenstates for more than four nucleons.
That is a challenging problem, however,
since it requires the formulation of a nucleon-pair shell model~\cite{Chen93,Chen97}
in an isospin-invariant formalism. 
In this paper we choose to extend our analysis
toward higher hole number
through the boson mapping techniques
explained in Sect.~\ref{s_map}.
It is important to stress that this approximation
goes beyond the original proposal of Blomqvist
since it involves an additional assumption
of the boson character of the fermion pairs.
The results presented in this subsection
therefore do not directly address Blomqvist's conjecture.

Once the mapping is carried out for the two- and four-hole systems
according to one of the two procedures described in Sect.~\ref{s_map},
a Hamiltonian is obtained in terms of the selected bosons $\ell_1,\ell_2,\dots,\ell_p$,
which can then be applied to systems with two or more bosons.
Such a description shall be referred to as $\ell_1\ell_2\dots\ell_p$-IBM,
where IBM stands for interacting boson model~\cite{Iachello87}.
Note that all versions of IBM thus obtained are isospin invariant;
for example, the $sd$-IBM is in fact the IBM-3 of Elliott and White~\cite{Elliott80}.

To compare the merits of different selections of fermion pairs,
the following mappings are considered:
\begin{enumerate}
\item
A single fermion pair $B$ with $J=9,T=0$, leading to the $b$-IBM.
\item
Two fermion pairs $S$ and $B$ with $J=0,T=1$ and $J=9,T=0$,
leading to the $sb$-IBM.
\item
Two fermion pairs $S$ and $D$ with $J=0$ and 2, both with $T=1$,
leading to the $sd$-IBM.
\item
Three fermion pairs $S$, $D$, and $G$ with $J=0$, 2, and 4, all with $T=1$,
leading to the $sdg$-IBM.
\end{enumerate}
The first two cases are inspired by Blomqvist's conjecture,
involving the aligned n-p pair,
while the next two are the standard choice of the IBM~\cite{Iachello87}
and its most frequently used extension which includes $g$ bosons.
(For a review on the latter, see Ref.~\cite{Devi92}.) 

\begin{table}
\caption{
Energies (in MeV) of $T=0$ levels for four nucleons in the $1g_{9/2}$ orbit ($^{96}$Cd)
calculated with the shell-model interaction SLGT0
and compared with various versions of IBM
obtained by democratic or OAI mapping.
$E_0$ is the binding energy of the ground state.}
\label{t_4p} 
\begin{ruledtabular}
\begin{tabular}{rcccccccccccccccc}
&~~~&SLGT0&~~~&$b$-IBM&~~~&$sb$-IBM&~~~&$sd$-IBM&~~~&$sdg$-IBM\\
\hline
$E_0$     &&9.050&&8.643&&9.041&&8.932&&9.050\\
\hline
$0_1^+  $&&0       &&0       &&0       &&0       &&0\\
$2_1^+  $&&0.963&&0.678&&1.077&&1.199&&1.002\\
$4_1^+  $&&2.100&&1.941&&2.339&&3.754&&2.204\\
$6_1^+  $&&3.079&&3.302&&3.700&&---      &&4.034\\
$8_1^+  $&&3.449&&4.425&&4.824&&---      &&5.688\\
$10_1^+$&&5.227&&5.179&&5.578&&---       &&---\\
$12_1^+$&&5.904&&5.572&&5.971&&---      &&---\\
$14_1^+$&&6.056&&5.692&&6.091&&---      &&---\\
$16_1^+$&&5.904&&5.496&&5.895&&---      &&---\\
$18_1^+$&&---     &&$\infty$&&$\infty$&&--- &&---\\
$0_2^+  $&&4.594&&---     &&4.613&&4.491 &&4.594\\
$2_2^+  $&&4.491&&---     &&---      &&4.730 &&4.554\\
$4_2^+  $&&4.390&&---     &&---      &&---      &&4.538\\
\end{tabular}
\end{ruledtabular}
\end{table}
The results obtained with the various boson Hamiltonians
are compared with $T=0$ eigenstates of the SLGT0 interaction
for four, six, and eight nucleons
in Tables~\ref{t_4p}, \ref{t_6p}, and \ref{t_8p}, respectively.
The numerical calculations have been performed
with the codes {\tt ArbModel}~\cite{Heinze} and {\tt IBM-3}~\cite{Isacker}.
The former is a general purpose program
that can handle systems of fermions and/or bosons with arbitrary spins
and can thus be used for the shell-model as well as the IBM calculations;
the latter code is specifically written for the isospin-invariant $sd$-IBM.
Alternatively, for three and four identical bosons ({\it i.e.}, for $b$-IBM)
the calculations can be performed
with the expressions given in Sect.~\ref{s_4p}
and equivalent ones for the three-hole case.

A few remarks are in order.
All results concern {\em absolute} energies.
In the first line of each table are given the binding energies $E_0$
of the $T=0$ ground states, as obtained in the various mappings,
which should be compared with the corresponding quantity in the shell model.
In subsequent lines are given the energies of a selected number of states,
relative to this ground state.
This might lead to some seemingly counterintuitive results.
For example, it is seen from Table~\ref{t_g9}
that the four-hole $2^+_1$ shell-model state
overlaps 97~\% with a $B^2$ configuration.
Why, then, should its excitation energy come out rather poorly in $b$-IBM
(0.678~MeV compared with 0.963~MeV in the shell model, see Table~\ref{t_4p})?
The reason is that the absolute energy of the $2^+_1$ state
is rather well reproduced
(it misses only 0.122~MeV of the shell-model correlation energy)
while the absolute energy of the ground state
is rather worse underbound (by 0.407~MeV) in $b$-IBM.

\begin{table}
\caption{
Energies (in MeV) of $T=0$ levels for six nucleons in the $1g_{9/2}$ orbit ($^{94}$Ag)
calculated with the shell-model interaction SLGT0
and compared with various versions of the IBM
obtained with two methods of mapping, democratic (Dem) and OAI.
$E_0$ is the binding energy of the ground state.}
\label{t_6p}
\centering
\begin{tabular}{ccccccccccccccccccc}
\hline\hline
&~~&SLGT0&~~~&$b$-IBM
&~~~&\multicolumn{3}{c}{$sb$-IBM}
&~~~&\multicolumn{3}{c}{$sd$-IBM}\\
\cline{7-9}\cline{11-13}
&&&&&&Dem&&OAI&&Dem&&OAI\\
\hline
$E_0$     &&11.276&&11.368&&11.368&&11.368&&8.592&&8.592\\
\hline
$0_1^+$  &&---      &&---     &&---     &&---      &&---     &&---\\
$1_1^+$  &&0.126&&4.340&&4.340&&4.340&&0       &&0\\
$2_1^+$  &&1.580&&---      &&---     &&---     &&---      &&---\\
$3_1^+$  &&0.298&&3.540&&3.540&&3.540&&0.547&&0.547\\
$4_1^+$  &&1.531&&3.848&&3.848&&3.848&&---      &&---\\
$5_1^+$  &&0.674&&2.163&&2.163&&2.163&&---      &&---\\
$6_1^+$  &&1.354&&2.352&&2.352&&2.352&&---      &&---\\
$7_1^+$  &&0      &&0        &&0       &&0      &&---      &&---\\
$8_1^+$  &&0.380&&0.505&&0.505&&0.505&&---      &&---\\
$9_1^+$  &&0.432&&0.796&&0.572&&0.538&&---      &&---\\
$10_1^+$&&1.579&&1.784&&1.784&&1.784&&---      &&---\\
$11_1^+$&&1.572&&1.833&&1.833&&1.833&&---      &&---\\
$12_1^+$&&2.933&&3.351&&3.351&&3.351&&---      &&---\\
$13_1^+$&&2.734&&3.220&&3.220&&3.220&&---      &&---\\
$14_1^+$&&3.840&&4.857&&4.857&&4.857&&---      &&---\\
$15_1^+$&&3.577&&4.602&&4.602&&4.602&&---      &&---\\
$16_1^+$&&5.364&&6.029&&6.029&&6.029&&---      &&---\\
$17_1^+$&&5.219&&5.677&&5.677&&5.677&&---      &&---\\
$18_1^+$&&6.606&&6.772&&6.772&&6.772&&---      &&---\\
$19_1^+$&&6.155&&6.324&&6.324&&6.324&&---      &&---\\
$20_1^+$&&---      &&$\infty$&&$\infty$&&$\infty$&&---      &&---\\
$21_1^+$&&6.464&&6.609&&6.609&&6.609&&---      &&---\\
\hline\hline
\end{tabular}
\end{table}
The two-boson spectra obtained
from mapping the four-hole shell-model results
depend on the kind of pairs included in the basis
but otherwise they are identical in the OAI and democratic mappings.
Therefore, for each of the different IBM versions,
there is a unique spectrum shown in Table~\ref{t_4p}
which is identical to that of the shell-model hamiltonian
when diagonalized in the corresponding (possibly truncated) two-pair basis.
While the OAI and democratic mappings
yield the same energy spectrum for four holes,
they lead to different boson-boson interaction matrix elements.
Hence, the OAI and democratic results
are different for the six- and eight-hole spectra
shown in Tables~\ref{t_6p} and~\ref{t_8p}.

If the number of two-pair states
equals the number of independent four-hole shell-model states,
then the two-boson calculation reproduces the four-hole results exactly.
According to Table~\ref{t_4p} this happens, for example,
for the three $J^\pi=0^+$ states with $T=0$
which can be exactly described as combinations of
$|S^2;0\rangle$, $|D^2;0\rangle$, and $|G^2;0\rangle$.
Consequently, the three shell-model $0^+$ states
are exactly reproduced in $sdg$-IBM.

Because of the Pauli exclusion principle,
no four-hole shell-model state exists with $J=18$
while this angular momentum is allowed
in the coupling of two bosons with $J=9$.
This is an example of the complication mentioned at the end of Sect.~\ref{s_map}
and the solution given there should be applied.
In this case it implies that the matrix element
$\langle B^2;18|\hat H^{\rm B}_2|B^2;18\rangle$
should be taken infinitely repulsive
and it is only by adhering to this procedure
that reasonable results can be obtained.

Not much is known experimentally about $^{94}$Ag 
except for the presence of two isomers,
with tentative spin-parity assignments $7^+$ (presumably the lowest $T=0$ state)
and $21^+$ at about $5.78(3)$~MeV above the $7^+$~\cite{Mukha05}.
The different shell-model interactions SLGT0, GF, and Nb90
all predict a $7^+$ as the $T=0$ ground state,
and a $21^+$ level at 6.464, 5.948, and 4.632~MeV, respectively.
The $b$-IBM reproduces the shell-model result
for the binding energies of these isomers
to about 100~keV for the $7^+$
and less than that for the $21^+$
(see Table~\ref{t_6p} for the results of the SLGT0 interaction).
This result is valid for the different shell-model interactions:
although the binding energies calculated with the three shell-model interactions
vary by several MeV,
in each case they are matched
by the (appropriately mapped) $b$-IBM to within about 100~keV,
indicating that the $B$ pair incorporates
most of the correlations for the $7^+$ and $21^+$ states.

The $b$-IBM results should be contrasted to those obtained with $sd$-IBM
which fails completely to reproduce the spectroscopy of $^{94}$Ag.
This is not surprising since it is known
from the work of Elliott and Evans~\cite{Elliott81}
that IBM-3 cannot give a satisfactory description of odd-odd nuclei
which require the addition of isoscalar $s$ and $d$ bosons
leading to the so-called IBM-4.
While the latter is a realistic model
when low-$j$ shell-model orbits are involved
({\it e.g.}, for $sd$-shell nuclei~\cite{Halse84,Halse85}),
the present results seem to indicate
that the mapping from a shell-model space with high-$j$ orbits
calls for the inclusion of an aligned isoscalar n-p pair with $J=2j$.

\begin{table}
\caption{
Energies (in MeV) of $T=0$ levels for eight nucleons in the $1g_{9/2}$ orbit ($^{92}$Pd)
calculated with the shell-model interaction SLGT0
and compared with various versions of the IBM
obtained with two methods of mapping, democratic (Dem) and OAI.
$E_0$ is the binding energy of the ground state.}
\label{t_8p}
\centering
\begin{tabular}{ccccccccccccccccccc}
\hline\hline
&~~&SLGT0&~~~&$b$-IBM
&~~~&\multicolumn{3}{c}{$sb$-IBM}
&~~~&\multicolumn{3}{c}{$sd$-IBM}\\
\cline{7-9}\cline{11-13}
&&&&&&Dem&&OAI&&Dem&&OAI\\
\hline
$E_0$     &&18.937&&18.135&&18.771&&18.646&&18.624&&19.999\\
\hline
$0_1^+$  &&0       &&0       &&0       &&0       &&0       &&0\\
$2_1^+$  &&0.927&&0.637&&1.170&&0.917&&0.728&&0.762\\
$4_1^+$  &&1.728&&1.104&&1.740&&1.608&&1.561&&2.054\\
$6_1^+$  &&2.512&&1.965&&2.628&&2.441&&3.155&&4.267\\
$8_1^+$  &&3.198&&2.836&&3.501&&3.320&&5.486&&6.861\\
$10_1^+$&&4.233&&3.683&&4.325&&4.185&&---      &&---\\
$12_1^+$&&5.123&&4.414&&5.050&&4.924&&---      &&---\\
%$14_1^+$&&4.332&&3.747&&?        &&?       &&---      &&---\\
%$16_1^+$&&4.634&&4.240&&?        &&?       &&---      &&---\\
\hline\hline
\end{tabular}
\end{table}
While $b$-IBM and $sb$-IBM are largely equivalent
for the odd-odd nucleus $^{94}$Ag,
this is not the case for the even-even nucleus $^{92}$Pd.
As can be seen from Table~\ref{t_8p},
the $s$ boson provides crucial additional correlation energy
which brings the boson result close to its shell-model equivalent.
This nucleus was studied recently in a fusion-evaporation experiment~\cite{Cederwall10}.
The excitation energies of the yrast levels,
calculated with the SLGT0 interaction,
are in reasonable agreement with the observed values of 0.874, 1.786, and 2.535~MeV, respectively.

\subsection{Electric quadrupole properties}
\label{ss_e2}
A further test of the aligned-n-p-pair hypothesis
can be obtained from electric quadrupole (E2) transition properties.
The E2 operator in the shell model is given by
\begin{equation}
\hat T^{\rm F}_\mu({\rm E}2)=
e_\nu\sum_{i\in\nu}r_i^2 Y_{2\mu}(\theta_i,\phi_i)+
e_\pi\sum_{i\in\pi}r_i^2 Y_{2\mu}(\theta_i,\phi_i),
\label{e_te2f1}
\end{equation}
where the sums are over neutrons and protons,
and each sum is multiplied with the appropriate effective charge.
This operator can be written alternatively
as a sum of an isoscalar operator, multiplied by $(e_\nu+e_\pi)$,
and an isovector operator, multiplied by $(e_\nu-e_\pi)$.
For the E2 transitions between $T=0$ levels of interest here,
only the former part contributes.
In second quantization,
which is a convenient formalism for carrying out the mapping,
the fermion E2 operator can be written as
\begin{equation}
\hat T^{\rm F}_\mu({\rm E}2)=
-\sqrt{\frac{55}{3\pi}}l_{\rm ho}^2
\left[e_\nu(a^\dag_\nu\times\tilde a_\nu)^{(2)}_\mu+
e_\pi(a^\dag_\pi\times\tilde a_\pi)^{(2)}_\mu\right],
\label{e_te2f2}
\end{equation}
where $a^\dag_\rho$ creates a neutron ($\rho=\nu$) or a proton ($\rho=\pi$)
in the $1g_{9/2}$ orbit,
and $\tilde a_{jm}=(-)^{j+m}a_{j-m}$.
Furthermore, the factor in front
comes from the radial integral over harmonic-oscillator wave functions
(with length parameter $l_{\rm ho}$)
involving the $1g_{9/2}$ orbit.

The lowest-order bosonic image of the fermion E2 operator 
is defined by the diagonal (reduced) matrix element
in the $9^+$ state of the 1n-1p system
which is given by
\begin{equation}
\langle(1g_{9/2})^2;9^+||\hat T^{\rm F}({\rm E}2)||(1g_{9/2})^2;9^+\rangle=
-\sqrt{\frac{55}{3\pi}}l_{\rm ho}^2\times\sqrt{\frac{1330}{187}}(e_\nu+e_\pi).
\label{e_rme2f}
\end{equation}
The E2 operator of the $b$-IBM
is of the form
\begin{equation}
\hat T^{\rm B}_\mu({\rm E}2)=
e_{\rm b}(b^\dag\times\tilde b)^{(2)}_\mu,
\label{e_te2b}
\end{equation}
and is necessarily of scalar character in isospin.
Since the mapping implies the equality
\begin{equation}
\langle(1g_{9/2})^2;9^+||\hat T^{\rm F}({\rm E}2)||(1g_{9/2})^2;9^+\rangle=
\langle b||\hat T^{\rm B}({\rm E}2)||b\rangle,
\label{e_rme2b}
\end{equation}
and the since the boson matrix element on the right-hand side
equals $\sqrt{5}e_{\rm b}$,
we find the following expression of the boson effective charge $e_{\rm b}$
in terms of the shell-model neutron and proton effective charges:
\begin{equation}
e_{\rm b}=-\sqrt{\frac{55}{3\pi}}l_{\rm ho}^2\times\sqrt{\frac{266}{187}}(e_\nu+e_\pi).
\label{e_eb}
\end{equation}
In the following, the factor $\sqrt{55/3\pi}(e_\nu+e_\pi)l_{\rm ho}^2$
is divided out of all matrix elements, fermionic as well bosonic.

\begin{figure}
\includegraphics[width=8.5cm]{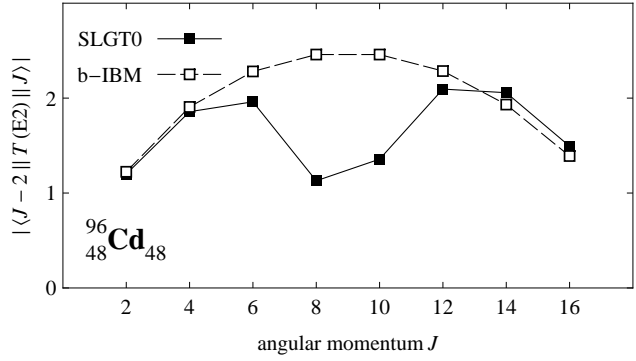}
\caption{
\label{f_e2cd96}
Absolute values of the E2 reduced matrix elements
for the transitions $J\rightarrow J-2$
between four-nucleon-hole states with $T=0$ in the $1g_{9/2}$ orbit ($^{96}$Cd),
calculated with the SLGT0 shell-model interaction
and compared with the mapped $b$-IBM.
Matrix elements are expressed in units $\sqrt{55/3\pi}(e_\nu+e_\pi)l_{\rm ho}^2$
(see text).}
\end{figure}
A first test is from E2 transitions between four-nucleon-hole states with $T=0$.
The shell-model results obtained with the SLGT0 interaction,
shown in Fig.~\ref{f_e2cd96},
display a characteristic decrease in quadrupole strength for $J\approx8$
which can be viewed as a remnant of a seniority-like classification.
The figure also shows the results found in $b$-IBM
using the boson effective charge derived in Eq.~(\ref{e_eb}),
with no adjustable parameter.
Not surprisingly, given that the $J=8$ state is poorly described
in terms of aligned n-p pairs (and hence $b$ bosons),
the two transitions involving this state deviate strongly in $b$-IBM
from the corresponding shell-model result.
Other transitions agree, however.

\begin{figure}
\includegraphics[width=8.5cm]{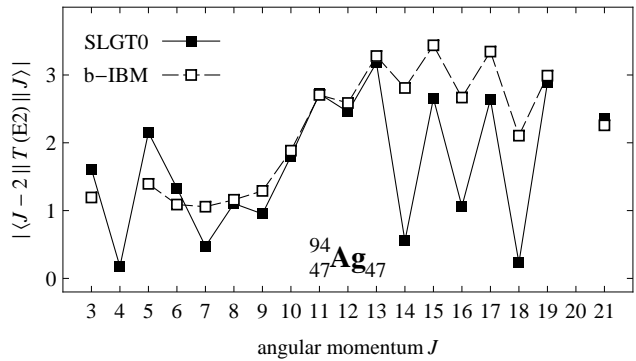}
\caption{
\label{f_e2ag94a}
Absolute values of the E2 reduced matrix elements
for the transitions $J\rightarrow J-2$
between six-nucleon-hole states with $T=0$ in the $1g_{9/2}$ orbit ($^{94}$Ag),
calculated with the SLGT0 shell-model interaction
and compared with the mapped $b$-IBM.
Matrix elements are expressed in units $\sqrt{55/3\pi}(e_\nu+e_\pi)l_{\rm ho}^2$
(see text).}
\end{figure}
\begin{figure}
\includegraphics[width=8.5cm]{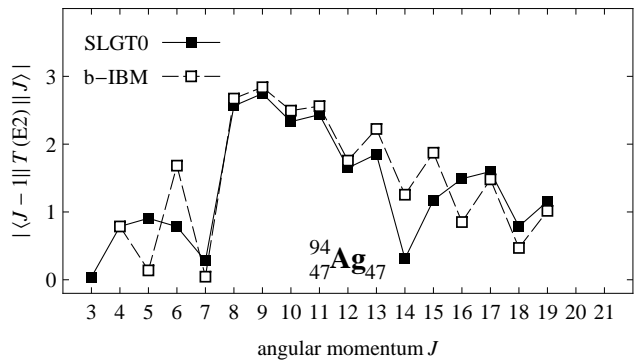}
\caption{
\label{f_e2ag94b}
Same caption as Fig.~\ref{f_e2ag94a} for $J\rightarrow J-1$ transitions.}
\end{figure}
A second test is provided by E2 transitions between six-nucleon-hole states with $T=0$.
They are shown in Figs.~\ref{f_e2ag94a} and \ref{f_e2ag94b}
for $J\rightarrow J-2$ and $J\rightarrow J-1$, respectively.
Assuming that an agreement between shell model and $b$-IBM is obtained only
if both the initial and final states are adequately represented by $b$ bosons,
we conclude that the $b$-IBM is a good approximation
for two ranges of angular momenta, namely $J=6$ to 13 and $J=17$ to 21.
This conclusion agrees, at least qualitatively, with the one drawn
on the basis of energies (see Table~\ref{t_8p}).
Since three $b$ bosons cannot couple to total angular momentum $J=2$,
this state is absent from $b$-IBM
while it is present in the shell model (see Table~\ref{t_8p}).
As a consequence, no $4\rightarrow2$ or $3\rightarrow2$ transitions occur in $b$-IBM
(see Figs.~\ref{f_e2ag94a} and \ref{f_e2ag94b}).
These transitions exist in the shell model
but it is rather striking that the calculated matrix elements
are very small indeed.

\begin{figure}
\includegraphics[width=8.5cm]{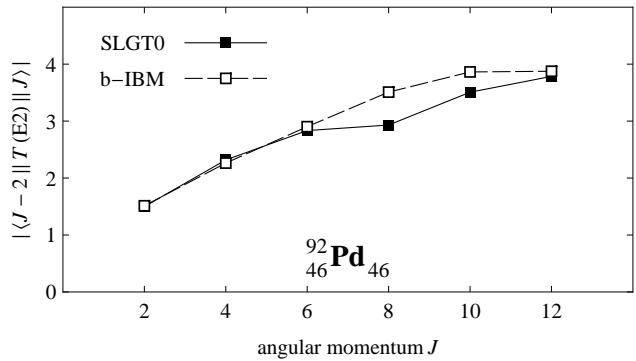}
\caption{
\label{f_e2pd92}
Same caption as Fig.~\ref{f_e2cd96}
for eight-nucleon-hole states ($^{92}$Pd).}
\end{figure}
Finally, in Fig.~\ref{f_e2pd92} are shown
the E2 transitions between eight-nucleon-hole states with $T=0$.
A small depletion of the E2 strength calculated in the shell model
is perceptible for $J\approx8$ and is absent in $b$-IBM.
Apart from this deviation both calculations agree,
indicating that the shell-model wave functions
can be adequately represented in terms of a single $b$ boson.
We emphasize once more that,
the number of states
that can be written in terms of $B$ pairs is small (of the order of 10),
no effective charges are needed to arrive
at the agreement in Fig.~\ref{f_e2pd92}.

Let us now formulate an overall evaluation of these results.
While the $b$-IBM gives in all cases an reasonable description
of the ground-state binding energy,
the addition of the standard $s$ boson (with $J=0,T=1$)
further improves the agreement.
In fact, the energies obtained in $sb$-IBM
(both with the OAI and democratic mappings)
agree well with those of the shell-model levels except for
(i) the $J=8$ level in the four-hole system,
(ii) low-$J$ levels of the six-hole system,
and (iii) levels with $J=14$ to 16 in the six-hole system.
The first discrepancy is obviously related
to the small overlap of the $J=8$ shell-model state with $|B^2;J=8\rangle$,
noted in Table~\ref{t_g9} and explained in Sect.~\ref{ss_sma}.
The second difference is also understandable
since a correct description of the low-$J$ states
in the odd-odd nucleus $^{94}$Ag
requires the consideration of low-$J$ pairs with $T=0$
which have been omitted from the present mapping.
The third deviation could be related to an unfavorable coupling
of three $B$ pairs to the angular momenta $J=14$, 15, and 16,
akin to the coupling of two $B$ pairs to $J=8$.
The results as regards E2 transitions
are consistent with what is concluded from the analysis of spectra.

\section{Conclusions}
\label{s_conc}
We have shown in this paper that
part of the low-energy spectroscopy of $N=Z$ nuclei
which have their valence nucleons confined to a single high-$j$ orbit,
can be represented in terms of an aligned isoscalar n-p $B$ pair with $J=2j$
and is further improved by the inclusion of an isovector $S$ pair coupled to $J=0$.
This was proven explicitly for a four-hole system
and indirectly, through a mapping onto a corresponding boson system,
for six and eight holes.
Some deficiencies were found in this approach.
A first concerns states of the four-hole system
with angular momentum $J\approx2j$
which turn out to be poorly approximated with just $S$ and $B$ pairs.
A second deficiency is more generic
and pertains to the low-$J$  states in odd-odd $N=Z$ nuclei,
the description of which calls for the inclusion
of isoscalar n-p pairs with low angular momentum.
Nevertheless, it should be noted that the two isomers
that have been observed so far in $^{94}$Ag, $(7^+)$ and $(21^+)$,
are adequately described in terms of $b$ bosons.

These results were obtained for the $1g_{9/2}$ orbit
and for three different choices of two-body interaction.
To what extent are they valid in general
and can they be considered as representative
of a system of neutrons and protons confined to a high-$j$ orbit?
In essence, two ingredients, geometry and dynamics,
determine the outcome of the present pair analysis.
The geometry is defined by the CFPs
and, provided $j$ is not too small,
is expected to evolve only slowly with $j$.
(It would in fact be an exercise of some interest
to perform the pair analysis in the limit of large $j$.)
The dynamics is determined by the two-body interaction
which in our study was varied significantly but within reasonable bounds.
The matrix elements shown in Table~\ref{t_me}
are typical of what is obtained for a residual interaction
with a short-range, attractive character~\cite{Schiffer76}
and we may thus expect similar results
when we move to orbits other than $1g_{9/2}$.

This work calls for further studies.
The pair analysis of the shell-model wave functions
should be extended to higher particle numbers
which can be achieved through an isospin-invariant formulation
of the nucleon-pair shell model.
Consequently, the present results require further confirmation
at higher particle number
but one is tempted to conclude at this point
that a realistic model can be formulated in terms of $s$ and $b$ bosons.
Due to its simplicity, such a model could be of use to elucidate
the main structural features of $N\sim Z$ nuclei in this mass region.
These topics are currently under study.

We wish to thank Stefan Heinze
for his help with the numerical calculations with {\tt ArbModel}
and Aurore Dijon for her help with the shell-model calculations.
This work has been carried out in the framework of CNRS/DEF project N 19848.
S.Z.\ thanks the Algerian Ministry of High Education and Scientific Research for financial support.
This work was also partially supported by the
Agence Nationale de Recherche, France,
under contract nr ANR-07-BLAN-0256-03.

\end{document}